\documentclass[journal,transmag]{IEEEtran}

\usepackage{cite}
\usepackage{graphicx}
\usepackage{amssymb,amsmath}

\begin{document}

\title{Coherent magnetization dynamics in strongly quenched Ni thin films}


\author{
    \IEEEauthorblockN{Akira Lentfert\IEEEauthorrefmark{1}, Anulekha De\IEEEauthorrefmark{1}, Laura Scheuer\IEEEauthorrefmark{1}, Benjamin Stadtmüller\IEEEauthorrefmark{1,2}, Georg von Freymann\IEEEauthorrefmark{1,3}, }
    \IEEEauthorblockN{Martin Aeschlimann\IEEEauthorrefmark{1}, and Philipp Pirro\IEEEauthorrefmark{1}}

    \IEEEauthorblockA{\IEEEauthorrefmark{1}Department of Physics and Research Centre OPTIMAS, Rheinland-Pfälzische Technische Universität Kaiserslautern-}
    \IEEEauthorblockA{Landau, 67663 Kaiserslautern, Germany}
	\IEEEauthorblockA{\IEEEauthorrefmark{2}Institute of Physics, Johannes Gutenberg University Mainz, 55128 Mainz, Germany}
 \IEEEauthorblockA{\IEEEauthorrefmark{3}Fraunhofer Institute for Industrial Mathematics ITWM, 67663 Kaiserslautern, Germany}
}

\IEEEtitleabstractindextext{%
\begin{abstract}
The remagnetization process after ultrafast demagnetization can be described by relaxation mechanisms between the spin, electron, and lattice reservoirs. Thereby, collective spin excitations in form of spin waves and their angular momentum transfer play an important role on the longer timescales. In this work, we address the question whether the strength of demagnetization affects the coherency and the phase of the excited spin waves. We present a study of coherent magnetization dynamics in thin nickel films after ultrafast demagnetization using the all-optical, time-resolved magneto-optical Kerr-effect (tr-MOKE) technique. The largest coherent oscillation amplitude was observed for strongly quenched systems, showing the conservation of coherency for demagnetizations of up to 90\%. Moreover, the phase of the excited spin-waves increases with pump power, indicating a delayed start of the precession during the remagnetization. 
\end{abstract}

}

\maketitle

\pagestyle{empty}
\thispagestyle{empty}

\IEEEpeerreviewmaketitle

\section{Introduction}

\IEEEPARstart{T}{he} laser-induced modulation of the spin system on ultrashort timescales is an important step towards time-efficient data storage and writing and therefore has been investigated over the past decades in various material systems \cite{Beaurepaire1996, Koopmans2000, Guidoni2022, Hohlfeld1997}. This ultrafast demagnetization process can be described by a phenomenological three-temperature model, which involves the interactions between the electron, lattice, and spin reservoirs. During the fast de- and remagnetization processes, the excitations of the spin system are dominated by hot electron dynamics and single particle excitations such as Stoner- excitations and Elliot-Yafet spin-flip processes \cite{Stiehl2022, Illg2013,Carva2011, Longa2007,Koopmans2009}. Therefore, the role of low-energy collective excitations such as spin waves is comparably unexplored on these timescales. Studies have revealed that spin waves become especially dominant during the slower remagnetization processes, visible after the fast remagnetization as the precession motion of the magnetization \cite{Mondal2018,vanKampen2002}. For smaller excitations, the oscillation and relaxation of the magnetization can be described by the Landau-Lifshitz-Gilbert equation \cite{Gilbert2004}, whereas for strongly quenched systems, non-linear spin-wave dynamics need to be taken into account. For materials with significant magneto-elastic coupling, it was also shown that the photo-induced strain in the film might have an impact on the dynamics \cite{Kim2012,shin_vomir_kim_van_jeong_kim_2022}.  

In our work, we investigate the influence of the demagnetization on the coherency and phase of the excited spin waves using the time-resolved magneto-optical Kerr-effect (tr-MOKE) effect. By applying a tilted field, the starting phase of the magnetization oscillation is well-defined between the pulses, allowing to trace the changes while sweeping the pump fluence \cite{Ju1999, vanKampen2002}. The oscillation of the magnetization during the remagnetization can be attributed to the excited coherent spin waves. Various properties of the remagnetization and oscillations, such as phase, amplitude and lifetime, are extracted and compared for different degrees of quenching of the magnetization to extract the influence on the coherency of the spin system.

\section{Sample and setup}

In this work, an 11 $\mathrm{nm}$ nickel (Ni) film, deposited on a MgO substrate using molecular beam epitaxy and capped with a 3 $\mathrm{nm}$ thick $\mathrm{Al_2O_3}$ and Au capping layer, is investigated.

\begin{figure}[ht]
    \centering
	\includegraphics[width=2in]{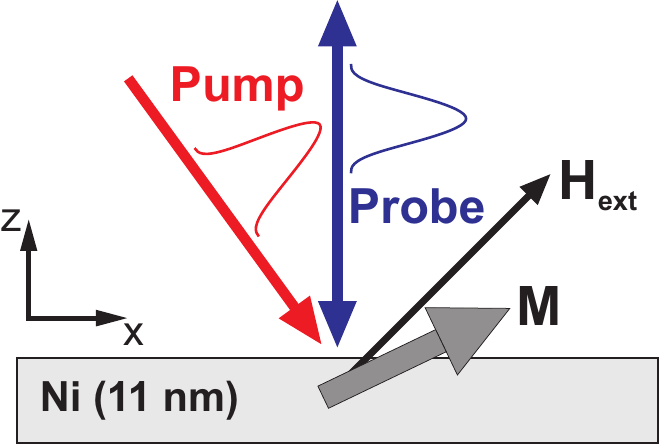}
	\caption{Geometry of the pump-probe experiments to detect the coherent oscillation after ultrafast demagnetization.}
	\label{Sample}
\end{figure}

The measurements are performed on a custom build setup based on the time-resolved magneto-optical Kerr-effect (tr-MOKE) with an amplified Ti:Sapphire laser system, where we pump with an intense, slightly defocused laser pulse at $\lambda =$ 800 $\mathrm{nm}$ with a pulse length of $t_{\mathrm{pump}} =$ 60 $\mathrm{fs}$ and a repetition rate of 1 $\mathrm{kHz}$. For the probe pulse, we chose the second harmonic with the same pulse duration to avoid effects like dichroic bleaching. To measure spin waves using MOKE, one needs to fix the phase of the precession motion for each pump pulse since the resulting signal is the average over multiple pulses. Therefore, the external magnetic field of $\mu_0H_{\mathrm{ext}} =$ 113 $\mathrm{mT}$ was applied under an angle of 45$^{\circ}$. As it was shown in \cite{Ju1999}, tilting the external field slightly out-of-plane triggers a reproducible and coherent precession of the magnetization and allows for the investigation of coherent spin waves after ultrafast demagnetization using MOKE. All the experiments were performed in polar geometry to detect the out-of-plane component of the magnetization, as shown in Fig. \ref{Sample}, since the induced changes of the in-plane component of the magnetization are too small to be detected compared to the static component.
After reflection on the sample, the probe beam passes through a lambda quarter plate to switch between the Kerr rotation and Kerr ellipticity signal, before it gets split into two orthogonal polarized beams by a Wollaston prism. These beams get both detected for both magnetic field directions to extract the change of the magneto-optic Kerr signal. In this work, the Kerr-ellipticity was recorded since the signal-to-noise ratio is superior to the Kerr-rotation signal.

\section{Results and discussions}

\begin{figure}[ht]
    \centering
	\includegraphics[width=3.5in]{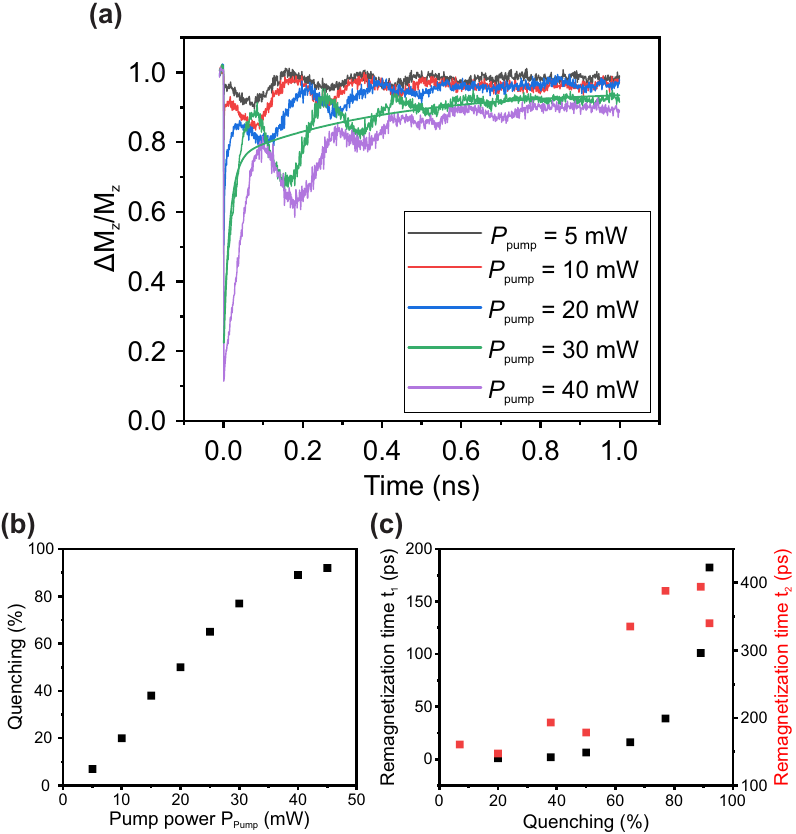}
	\caption{(a) The magnetization curves are normalized with respect to the initial magnetization level for selected pump powers. The curve for $P\mathrm{pump} =$ 30 $\mathrm{mW}$ is fitted with an exponential fit according to \ref{Exp_background}. (b) Dependence of the quenching of the magnetization on the pump power $P_\mathrm{pump}$. (c) Dependence of the remagnetization time $t_1$ on the quenching of the magnetization}
	\label{Ni_remag}
\end{figure}

Figure \ref{Ni_remag}(a) shows the magnetization traces for multiple pump powers $P_\mathrm{pump}$ in the 11 $\mathrm{nm}$ thick Ni film, where a quenching from a few percent up to almost full demagnetization of the sample was achieved. 

Furthermore, the remagnetization data were fitted with a two-phase exponential growth function, where the two time constants $t_1$ and $t_2$ correspond to the first and second remagnetization times, $y_0$ the offset and $A_1$ and $A_2$ represent the amplitude of the quenching:

\begin{equation}
    A_\mathrm{remag}(t) = A_1 e^{-t/t_1} + A_2 e^{-t/t_2} + y_0.
	\label{Exp_background}
\end{equation}

The first remagnetization part describes the relaxation of the phonon and electron system of the Ni layer into the substrate and the second remagnetization part describes the relaxation of the substrate. The quenching of the magnetization changes linearly for smaller $P_\mathrm{pump}$, as it is shown in Fig. \ref{Ni_remag} (b), and then saturates towards 100\% demagnetization, since the amount of additional energy to raise the temperature and at the same time decrease the magnetization further increases drastically. The huge amount of additional energy needed towards full demagnetization also leads to a longer remagnetization process, as can be seen in the asymptotic trend of $t_1$ and $t_2$ in Fig. \ref{Ni_remag} (c). The extracted remagnetization time $t_1$ and $t_2$ are plotted against the quenching of the demagnetization to make it comparable to other thin film measurements.

\begin{figure}[ht]
    \centering
	\includegraphics[width=3in]{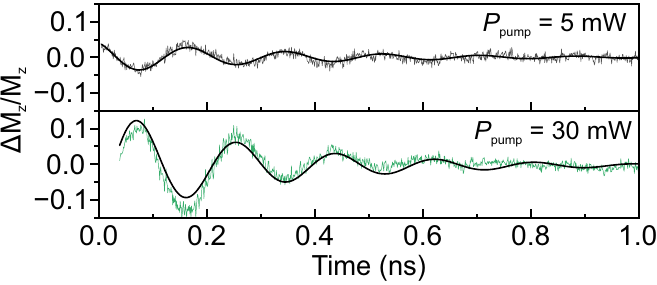}
	\caption{The resulting oscillation for $P\mathrm{pump} =$ 5 $\mathrm{mW}$ and $P\mathrm{pump} =$ 30 $\mathrm{mW}$ after removing the exp. background of the ultrafast de- and remagnetization, fitted with the damped sine function Eq.\ref{Sine_func}.}
	\label{Ni_sine}
\end{figure}

Now, the properties of the excited coherent oscillations, which correspond to the ferromagnetic resonance, can be extracted by fitting the remaining oscillations as shown in Fig. \ref{Ni_sine}(b) with a damped sine function:

\begin{equation}
    A_\mathrm{osc}(t) = A e^{-t/t_0} \sin(\omega t + \Delta\psi).
    \label{Sine_func}
\end{equation}

\begin{figure}[ht]
    \centering
	\includegraphics[width=3in]{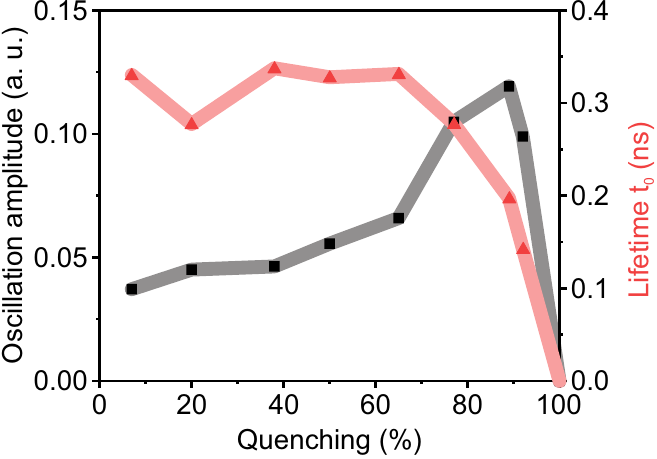}
	\caption{Oscillation amplitude and lifetime as a function of quenching of the magnetization, the lifetime is extracted from the damped sine fits.}
	\label{QuenchvsAmp}
\end{figure}

Here, $A$ represents the amplitude, $\omega$ the frequency, $\Delta\psi$ the phase shift and $t_0$ the lifetime of the oscillation. The extracted amplitude and lifetime are plotted against the quenching in Fig. \ref{QuenchvsAmp}. Here, the plotted amplitude is not the amplitude $A$ extracted from the fit of Eq. \ref{Sine_func}, but the real amplitude of the first maximum. Since the start of the oscillation and therefore the start of the fit is shifted in time, the amplitude taken from the damped sine fit would distort the actual oscillation amplitude.

For lower pump powers, the oscillation amplitude and lifetime remain rather constant. Further  increasing the pump fluence leads to an increase of the amplitude, whereas at the same time the lifetime starts to decrease. It is important to note that this does not mean that the lifetime of the spin waves is lowered, since with the tr-MOKE technique, only the coherent signal is measured.
When the quenching of the magnetization increases, the total amount of excited spin waves, including the coherent and incoherent contributions, increases, leading to a decrease of the effective magnetization. For pump powers larger than 30 $\mathrm{mW}$, the equilibrium magnetization is still not reached after 1 $\mathrm{ns}$ as it can be seen in Fig. \ref{Ni_remag} (a), indicating the presence of a large number of incoherent spin waves after the coherent oscillations vanish. The drop in the oscillation lifetime is therefore induced by non-linear multi-magnon scattering processes, where the spin waves are scattered to different states in the dispersion while conserving their energy and momentum. This scattering process is a threshold process which arises for large spin-wave amplitudes, and its dissipation rate is proportional to the magnon density of the involved spin-wave mode. Since these scattered spin waves will not appear in the coherent oscillations anymore, the lifetime of the oscillation is reduced for larger pump powers.
 
The largest amplitude of the coherent spin waves is observed for demagnetizations of about 90\%, which implies that the coherency of the spin system is still conserved for strongly quenched systems. Further demagnetization of the sample leads to a complete loss of magnetic order, and no coherent signal can be recorded with the tr-MOKE technique anymore. Nevertheless, it is expected that the total amount of spin waves increases continuously with increasing $P_\mathrm{pump}$.

\begin{figure}[ht]
    \centering
    \includegraphics[width=2.5in]{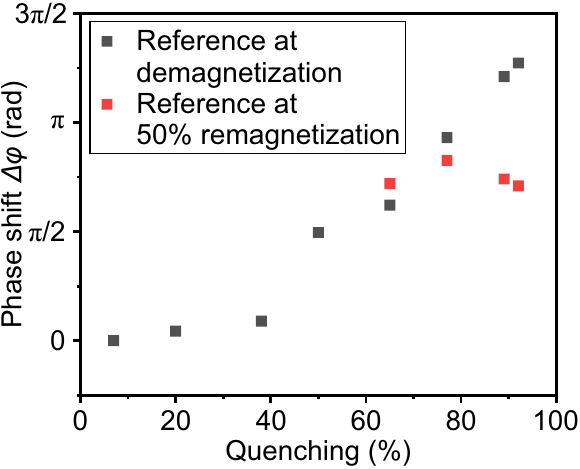}
    \caption{Phase shift $\Delta\psi$ of the coherent oscillations as a function of quenching of the magnetization for two different reference times.}
    \label{Time_shift}
\end{figure}

Next, the phase of the oscillation is plotted in relation to the quenching in Fig. \ref{Time_shift}. 

Although the pump pulse length is identical for all the measurements, the oscillations are delayed with increasing power. Since the demagnetization happens quasi-instantly on the time scale of the coherent precession for all pump powers, this delay must be caused during the remagnetization process. To underline this statement, the magnetization traces with a demagnetization larger than 50\% were analyzed in more detail. They were shifted in such a way, that the time zero is when the curves are remagnetized to 50\% and again, the phase of the oscillations was extracted.
Recently, it was  shown that the phase of the coherent oscillations after ultrafast demagnetization is strongly influenced by a quasi-static strain, where the fs-laser pulse leads to a thermal expansion of the lattice \cite{shin_vomir_kim_van_jeong_kim_2022}. By incorporating the magnetoelastic coupling into the Landau-Lifshitz-Gilbert model, the phase of the oscillations was reproduced for thicker metal layers. Hereby, the thermal effects tilt the effective field more out-of-plane towards the external field, whereas the quasi-static strain counteracts it, tilting the effective field more in-plane. Since nickel shows strong magnetoelastic coupling, the balance between these contributions might drastically change for thin films, depending on the applied pump fluence, leading to a phase change of the following coherent oscillation. In the same work, it is shown that the oscillations do not shift in time for a thicker nickel film (270~nm) on a SiO$_2$ substrate for various fluences. Together with the results of Fig. \ref{Time_shift}, one can conclude that the thermal effects have an impact on the magnetization dynamics during the remagnetization. Since MgO has a smaller heat conductivity than nickel, a significant change of the magnetization dynamics due to heating is possible \cite{Ho1974, Powell1966}. One more difference to their work is that the contribution of super-diffusive transport of hot electrons away from the surface during the demagnetization is smaller in the thin sample, leading to the changes we see here \cite{Battiato2010}.

Another possible explanation is that the resonance frequency of the system is non-linearly shifted due to the high-amplitude magnon excitation during the first remagnetization. Hereby, the frequency for spin waves in in-plane magnetized films is shifted downwards and therefore, the continuous change of the frequency during the first remagnetization would result in an increased final phase for stronger excitations. Another scenario that could explain this behavior is that due to the quasi-instant excitation of the system, the magnetization needs to remagnetize to a certain level to make linear coherent oscillations possible.

To investigate the influence of the nonlinear shift, one might study samples with a different sign of the nonlinearity, where the resonance condition is shifted to higher frequencies. Thus, with stronger demagnetization, the phase shift $\Delta\psi$ would decrease in this case. Furthermore, an additional heat sink in form of a thick gold layer could be added to increase the heat diffusion away from the pump spot to investigate the influence of heating on the remagnetization process. This would allow for the observation of the relation between the phase shift and the remagnetization time.

\section{Conclusion}
In this work, we analyzed the precession motion of the magnetization in a thin nickel film after strongly quenching the magnetization by pumping the system with an intense fs-laser pulse. We see that the coherency of the spin system is conserved, even in case that the magnetization is reduced by 90\%. At the same time, the lifetime of the coherent oscillation decreases due to multi-magnon scattering processes arising. Interestingly, the phase of the oscillations increases with pump fluence, indicating that the motion of the magnetization is delayed during the first remagnetization phase. Possible explanations are that (i) the quasi-static strain affects the effective field, changing the precession axis and influencing the phase of the oscillation, (ii) the thermal effects change the magnetic properties in nickel, leading to a change in the dynamics, (iii) the continuous non-linear shift of the resonance frequency due the large spin-wave amplitude leads to an accumulation of an additional phase during the first remagnetization and (iv) the magnetization needs to remagnetize to a certain level to make linear coherent oscillations possible.
\\

The authors acknowledge support by the Deutsche Forschungsgemeinschaft (DFG, German Research Foundation) through No. TRR 173-268565370 (Project B11).


\bibliographystyle{IEEEtran}
\bibliography{IEEEabrv}

\end{document}